\documentclass{elsart}

\usepackage{graphicx}
\usepackage{amsmath} % ams business is for use of 'cases'
\usepackage{amsfonts}
\usepackage{amssymb}
\usepackage{setspace}

\usepackage{amssymb}

\begin{document}

\begin{frontmatter}
\journal{Astroparticle Physics}



\title{Search for Correlations between Nearby AGNs and Ultra-high Energy Cosmic Rays}


\author[New Mexico]{J.~D.~Hague}
\author[New Mexico]{J.A.J.~Matthews\thanksref{email}}
\author[New Mexico]{B.~R.~Becker}
\author[New Mexico]{M.~S.~Gold}

\address[New Mexico]{University of New Mexico, Department of Physics
and Astronomy, Albuquerque, New Mexico, USA}

\thanks[email]{Corresponding author, E-mail:
\texttt{johnm@phys.unm.edu}}

\date{\today}
%\date{Revised: August 25, 2006}

\begin{abstract}
The majority of the highest energy cosmic rays are thought 
to be electrically charged: protons or nuclei.  Charged particles 
experience angular deflections as they pass through galactic 
and extra-galactic magnetic fields.  As a consequence correlation 
of cosmic ray arrival directions with potential sources has 
proved to be difficult.  This situation is not helped by current 
data samples where the number of cosmic rays/source are 
typically $\leq O(1)$.  Progress will be made when there are 
significantly larger data samples and perhaps with better 
catalogs of candidate sources.   This paper reports a search 
for correlations between the RXTE catalog of nearby active 
galactic nuclei, AGNs, and the published list of ultra-high 
energy cosmic rays from the AGASA experiment.  Although no 
statistically significant correlations were found, two 
correlations were observed between AGASA events and the most 
inclusive category of RXTE AGNs.
\end{abstract}


\end{frontmatter}

\section{Introduction}

\label{introduction}

Perhaps the primary goal of all experiments studying the highest energy cosmic
rays is to find the source of these particles.  While circumstantial
evidence may favor one type of source over another, 
demonstration of a clear correlation between the direction of cosmic 
rays and their sources is arguably essential.  Unfortunately for electrically 
charged cosmic rays, galactic magnetic fields, and for the
highest energy cosmic rays extra-galactic magnetic fields, 
cause angular deflections that can blur the correlation between 
cosmic ray arrival direction and source direction.  If the sources as viewed
from the earth are extended\,\cite{waxman,cuoco} the problem is even 
more difficult.  Unless otherwise noted, for this paper we assume 
compact (point-like) sources for the highest energy cosmic rays.  

If the angular blurring from magnetic fields is small\,\cite{dolag}
({\it i.e.} not significantly greater than the experimental angular
resolution) and/or for neutral primaries, then
experiments should observe cosmic rays that cluster in arrival 
direction\,\cite{agasa_cluster,tinyakov},
and/or that correlate with potential astronomical ({\it e.g.} BL Lac) 
sources\,\cite{tinyakov_BLLac,gorbunov_BLLac0,gorbunov_BLLac1,gorbunov_BLLac2,hires_BLLac}.  For nearby sources, where experiments should detect multiple
cosmic rays/source, event clusters provide bounds on the cosmic ray source 
density\,\cite{dubovsky,blasi,kachelriess} potentially 
favoring one type of source for the highest energy cosmic rays over another.
However at this time the situation is less than clear as some
results\,\cite{finley,hires_cluster} question the significance of the 
reported clusters and/or some of the BL Lac correlations\,\cite{hires_BLLac,not_BLLacs}.

If deflections of charged cosmic rays by extra-galactic magnetic 
fields are not small\,\cite{sigl}, then lower energy, $E$, cosmic rays
should experience the greatest angular deflections.  
Unfortunately small experiment data samples and a cosmic
ray flux $\propto E^{-3}$ have often caused studies to retain cosmic
rays to energies, $E_{thresh}$, well below 
GZK\,\cite{gzk} energies\,\cite{agasa_cluster}.
Furthermore deflections of the highest
energy cosmic rays even by our galactic magnetic field can be 
substantial\,\cite{tinyakov_Bfield,kachelriess_Bfield,tanco}.  
As magnetic deflections scale proportional to the charge of the
primary cosmic ray, nuclei in the cosmic rays may have significant
deflections.  Although most searches have looked for clustering and/or
source correlations on small angular scales, studies at larger angular
scales have also found evidence for clustering and/or
source correlations\,\cite{kachelriess_clusters,smialkowski,singh}.
Certainly the angular scale of cosmic ray clusters and 
the magnitude, and thus relevance, of the deflections of
ultra-high energy cosmic rays 
by magnetic fields is not universally agreed to at this time.

In the future, significantly larger data samples will allow analyses to
increase $E_{thresh}$ while retaining the number of observed cosmic 
rays/source (for nearby sources) $\geq O(1)$.   However another
possibility is to exploit catalogs of candidate sources.  With a catalog
of source directions, cosmic rays can be effectively correlated with
sources even when if magnetic field deflections are ``not small''
and/or when the the number of observed cosmic 
rays per source is $<1$ allowing searches with existing data samples.
That said, catalog based studies are limited by
the completeness of the source catalog and the relevance (or not)
of that class of astronomical source to the production of the highest
energy cosmic rays.  Often conjectured astrophysical sources include 
gamma ray bursts, GRBs, and/or active galactic nuclei, AGNs\,\cite{BGG2002}.

This paper reports a search for correlations between a catalog of nearby 
AGNs\,\cite{rxte_catalog} and the published list of ultra-high energy 
cosmic rays from AGASA\,\cite{agasa_cluster}.  The components of our
analysis are listed in Section\,\ref{section:components}.  Issues that
relate to data and AGN selection are given in 
Section\,\ref{section:selection}.  The cosmic ray--AGN comparison 
results are given in Section\,\ref{section:comparison}.  
Section\,\ref{section:summary} summarizes this study.

\section{Analysis Components}
\label{section:components}

Our comparison of ultra-high energy cosmic rays and a catalog of AGNs
includes three components: the RXTE catalog of AGNs, the AGASA list of 
cosmic rays, and a Monte Carlo sample of uniformly distributed cosmic rays 
generated to match the experimental acceptance of AGASA.

The catalog of nearby AGNs\,\cite{rxte_catalog} results from the
Rossi X-ray Timing Explorer, 
RXTE, all-sky slew survey\,\cite{rxte} sensitive to sources of
hard X-rays (3-20keV).  The survey excluded 
the galactic plane ($|b|>10^{\circ}$)
but covered $\sim 90$\% of the remaining sky.  X-ray sources were located to
better than $1^{\circ}$ and then correlated with known astronomical objects.
The efficiency for AGN identification was estimated to
be $\sim 70\%$ with somewhat higher efficiency for northern AGNs ($\sim 87\%$)
and somewhat lower efficiency for southern AGNs 
($\sim 60\%$)\,\cite{rxte_catalog}.  The resulting catalog provides 
source directions and probable source distances and intrinsic X-ray 
luminosities, $L_{3-20}$.  The catalog is best for nearby AGNs
as RXTE signal thresholds significantly reduced the 
efficiency for detecting distant sources; additional details are given below.

The list of ultra-high energy cosmic rays comes from published AGASA data 
\,\cite{agasa_cluster}.  

The Monte Carlo sample of uniformly distributed cosmic rays was generated
according to a $\cos(\theta)\sin(\theta)$ distribution in local zenith 
angle, $\theta\leq45^{\circ}$, and uniform in local azimuth.  
Events were then transformed to 
celestial right ascension and declination assuming constant detector 
aperture with time.
 
Correlations, between the AGASA events
and the catalog of AGNs from RXTE, would appear as an excess at small 
angular separations in comparison to the 
Monte Carlo sample of simulated cosmic rays.  To be clear,
define unit vectors in the directions of cosmic rays, \^{u}$_i$, 
AGNs, \^{v}$_j$, and Monte Carlo simulated cosmic rays, \^{w}$_k$.  
A correlation {\it signal} should then appear {\it near 1.0}
in the distribution of {\it dot}-products: ~\^{u}$_i \cdot $\^{v}$_j$~ 
(if magnetic field deflections are modest).  The index ``$i$'' runs over the 
cosmic rays in the data sample.  For each value of ``$i$'', only the
AGN catalog source (index ``$j$'') giving the maximum value 
of: ~\^{u}$_i \cdot $\^{v}$_j$~ contributes to the 
distribution\,\footnote{Thus 
each cosmic ray has one entry in the {\it dot}-product distribution.  This
choice is consistent with each cosmic ray having one source.  As only the
AGN source {\it nearest in angle} to the cosmic ray is chosen 
this can result in possible misidentification in the case of large 
source density.}.  The simulated distribution of
{\it random background} comes from the analogous distribution
of: ~\^{w}$_k \cdot $\^{v}$_j$ where index ``$k$'' now runs over the 
sample of Monte Carlo simulated cosmic rays.  As with the cosmic
ray events, only the
AGN catalog source (index ``$j$'') giving the maximum value 
of: ~\^{w}$_k \cdot $\^{v}$_j$~ contributes to the distribution.

\section{Cosmic Ray and AGN Selection}
\label{section:selection}

A few choices have been made in the comparison of AGASA data and
catalog of AGNs from RXTE.  These are described here.

\begin{figure}[h]
\centering
 \includegraphics[angle=-90, width=12cm]{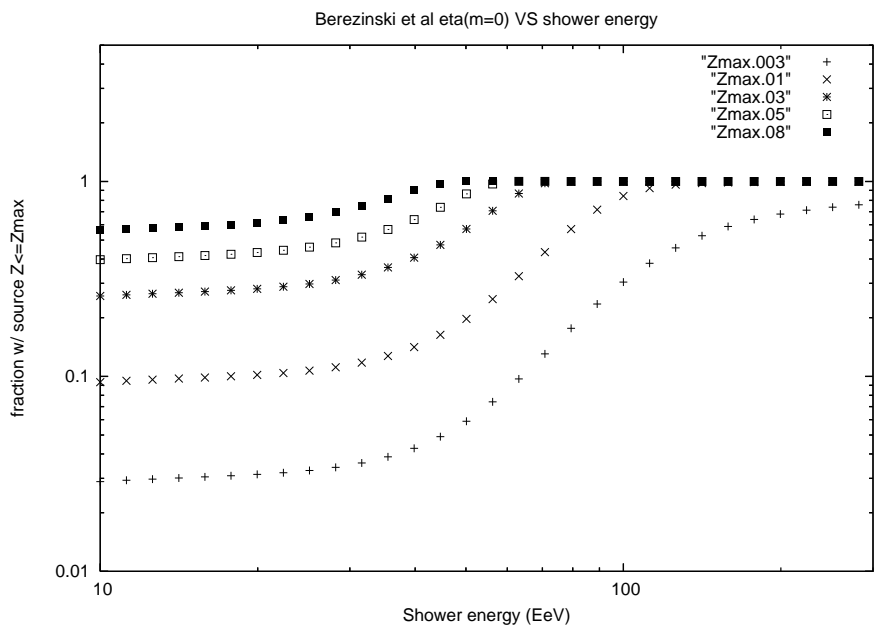}
 \caption{{\it \label{fig:berezinski}} The curves, from a GZK 
 model\,\cite{berezinski}, show the predicted fraction of cosmic
 ray events with source redshift, $z < z_{max}$
 versus cosmic ray energy for a selection of $z_{max}$ distances:
 $0.003 \leq z_{max} \leq 0.08$.    The GZK model assumed proton
 primaries with a power law spectrum at the source $\propto E^{-2.7}$.
 No enhancement factor of increasing source density with redshift, $z$,
 was included.  The result was derived from Fig.6 of Ref.\,\cite{berezinski}.}
\vskip 0.5 cm
\end{figure}

The AGASA data have energies, $E > 40$EeV and populate values of
declination: $-10^{\circ} \leq Dec \leq 80^{\circ}$.  As noted above,
the steep cosmic rays spectrum, $\propto E^{-3}$, and modest number of
events: 57 with $E>40$EeV and 29 (just over half) with $E>53EeV$ 
led us to consider three (overlapping) bins in 
energy: $E\geq40EeV$, $E\geq53EeV$ and $E\geq100$EeV.  The last was to see if
there are any correlations with the AGASA super-GZK events.
Except for the $E\geq100$EeV selection, most of the cosmic rays are
predicted, at least under the assumption 
of proton primaries\,\cite{berezinski},
to originate at values of redshift, $z>0.01$\,\footnote{This corresponds
to a distance $r \approx 42$ Mpc}; see Fig.\,\ref{fig:berezinski}.

\begin{figure}[h]
\centering
\includegraphics[angle=-90, width=12cm]{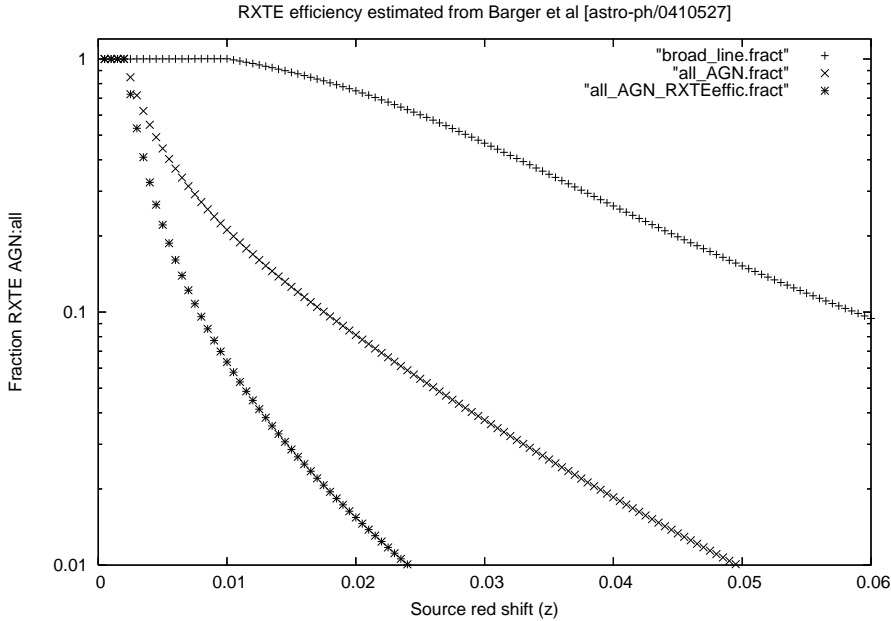}
 \caption{{\it \label{fig:AGNeffic}} The curves show the estimated
 RXTE source detection efficiency, {\it i.e.} the
 fraction of RXTE AGNs\,\cite{rxte_catalog} to all 
 AGNs\,\cite{barger}, {\it VS} source distance resulting from the RXTE 
 instrument detection threshold.  As noted in the text, 
 two categories (or definitions) of AGNs are considered: {\it all}-AGNs with 
 RXTE 3-20keV intrinsic luminosities, $L_{3-20} \geq 10^{41}$ ergs/s 
 shown as ``$\times$'',
 and {\it broadline}-AGNs with 2-8keV intrinsic luminosities, 
 $L_{2-8} \geq 10^{42}$ ergs/s shown as ``$+$''.  For redshift
 (distances) with full RXTE source detection efficiency then the fraction 
 is 1.  Parenthetically, we would obtain lower efficiencies 
 (for the {\it all}-AGN category)
 if we were to use the AGN number density {\it VS} X-ray luminosity 
 deduced by the RXTE experiment\,\cite{rxte_catalog}
 and shown as ``$*$''.}
\vskip 0.5  cm 
\end{figure}

To match the AGASA acceptance, we selected AGNs with 
$-10^{\circ} \leq Dec \leq 80^{\circ}$.  We have also made selections
on the redshift of the AGNs to consider sources only with RXTE source
detection efficiency $^>_{\sim}50$\%.  

The estimate of the RXTE source detection efficiency involves two issues:
\begin{enumerate}
\item the RXTE instrument source detection threshold ({\it i.e.} the
selection bias) {\it VS} redshift from Fig.1 of Ref.\,\cite{rxte_catalog},
\item the number density of AGNs {\it VS} redshift and 
intrinsic X-ray luminosity from Table 2 of Ref.\,\cite{barger}.
\end{enumerate}
Motivated by Ref.\,\cite{barger}, we divide the AGNs into the two
categories: {\it all}-AGNs and {\it broadline}-AGNs.  
For the {\it all}-AGN category we require that the X-ray 3-20keV 
intrinsic luminosity\,\footnote{For our study we relate the
RXTE 3-20keV intrinsic luminosities, $L_{3-20}$ in ergs/s, to 
2-8keV intrinsic luminosities, $L_{2-8}$ in ergs/s using: 
$L_{2-8} \approx L_{3-20}/2$; private communication from Sergey Sazonov.}, 
$L_{3-20} \geq 10^{41}$ ergs/s, to match the RXTE data.
With this intrinsic luminosity threshold 
the estimated {\it all}-AGN number density is $4.2 \times 10^{-4}$ Mpc$^{-3}$
consistent with the RXTE source density determination
of $\sim 5 \times 10^{-4}$ Mpc$^{-3}$\,\cite{rxte_catalog}.
For the {\it broadline}-AGN category we require that the X-ray
2-8keV intrinsic luminosity, $L_{2-8} \geq 10^{42}$ ergs/s as this
selects X-ray sources that are likely to be AGNs based purely on energetic 
grounds\,\cite{steffen}.  With this intrinsic luminosity threshold the 
estimated {\it broadline}-AGN number density is
$\sim 2 \times 10^{-5}$ Mpc$^{-3}$.

Combining the RXTE detection threshold with our definition
of two categories of AGN (above), we obtain the fraction of
each AGN category {\it VS} redshift.  This is shown in 
Fig.\,\ref{fig:AGNeffic}.  Based on this result we restrict the
redshifts for the {\it all}-AGN category to $z\leq0.005$ and
the redshifts for the {\it broadline}-AGN category to $z\leq0.03$.

\section{Cosmic Ray--AGN Comparisons}
\label{section:comparison}

\begin{figure}[h]
\begin{center}
\begin{tabular}{cc}
\includegraphics[width=0.5\linewidth]{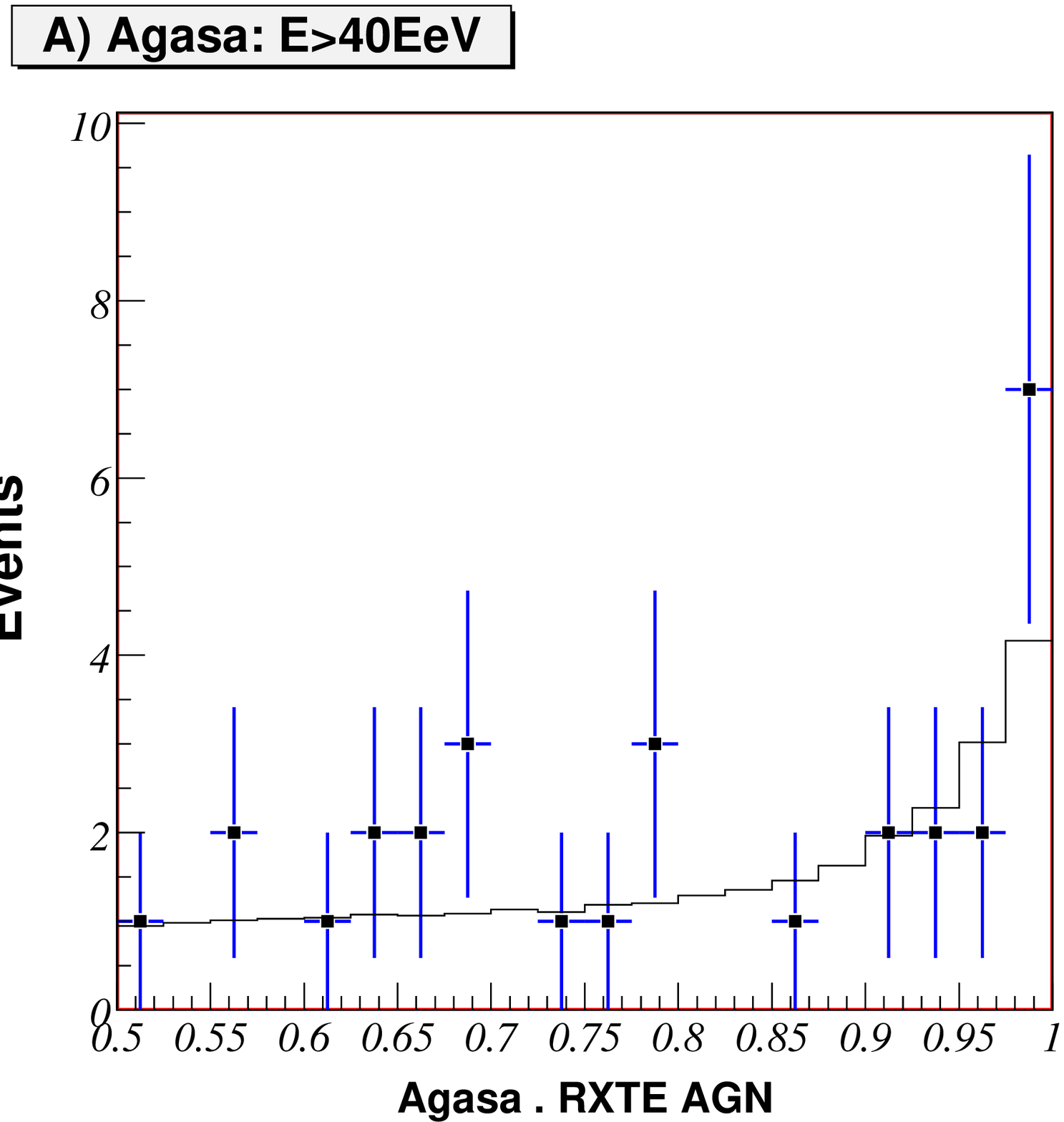} & 
\includegraphics[width=0.5\linewidth]{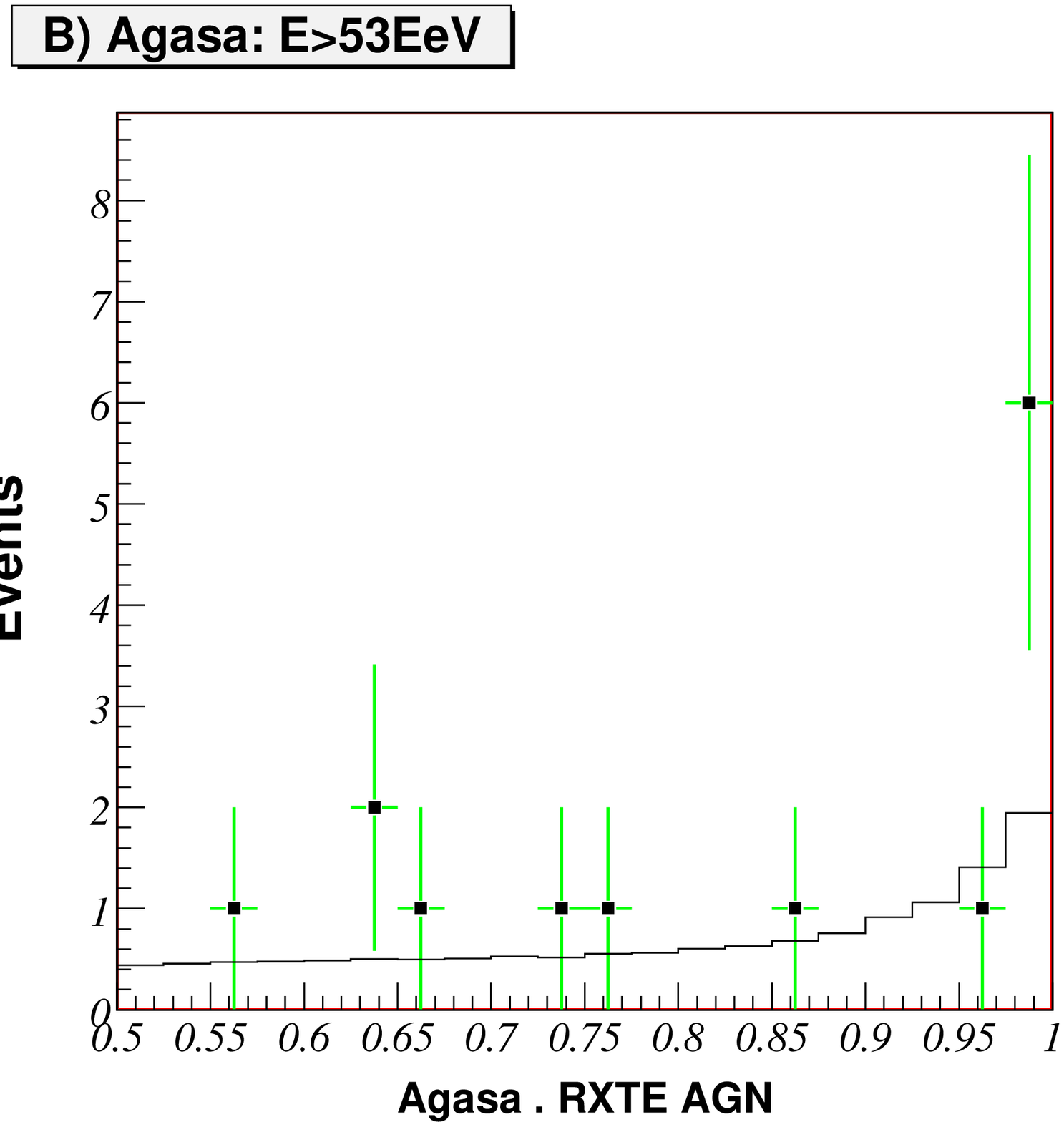}
\end{tabular}
\vspace{0.2cm}
\begin{tabular}{c}
\includegraphics[width=0.5\linewidth]{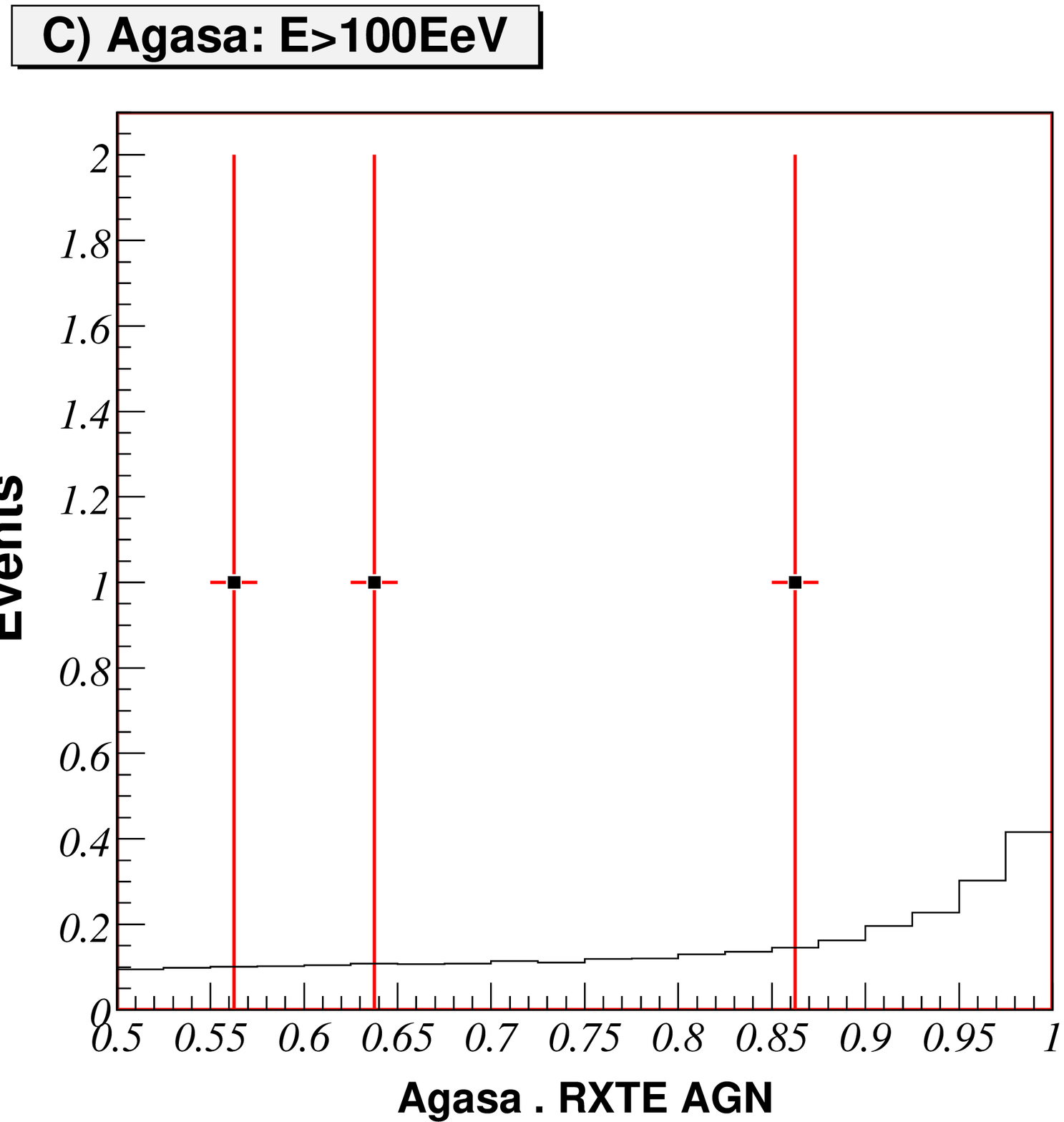}
\end{tabular}
\end{center}
 \caption{{\it \label{fig:allAGNplot}} The plots show the distribution
 of {\it dot}-products (see definition in the text)
 for the {\it all}-AGN selection: (top left) with
 cosmic ray energies $E \geq 40$EeV, (top right) with
 cosmic ray energies $E \geq 53$EeV, and (bottom) with
 cosmic ray energies $E \geq 100$EeV.  The curve on each figure shows
 the Monte Carlo {\it random background} normalized
 to the number of entries in each plot. }
\vskip 0.05 cm 
\end{figure}

\begin{figure}[h]
\centering
\includegraphics[width=14cm]{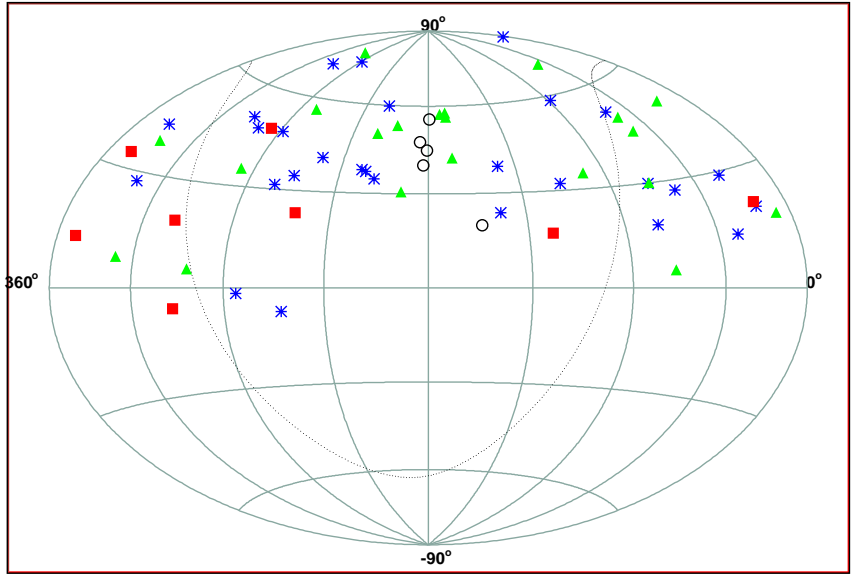}
 \caption{{\it \label{fig:allAGNdisplay}} The figure shows the 
 map of $RA-Dec$ for the AGASA data and the AGNs from 
 the {\it all}-AGN selection.  The AGASA data are plotted in
 {\it blue}($*$) for 40EeV$\leq E <$53EeV, 
{\it green}($\blacktriangledown$) for 53EeV$\leq E <$100EeV,
 and  {\it red}($\blacksquare$) 
for 100EeV$\leq E$.  The RXTE AGNs are plotted as {\it black}({\bf o}).  
The galactic plane is drawn as a dotted line. }
\vskip 0.2 cm 
\end{figure}

\begin{figure}[h]
\begin{center}
\begin{tabular}{cc}
\includegraphics[width=0.5\linewidth]{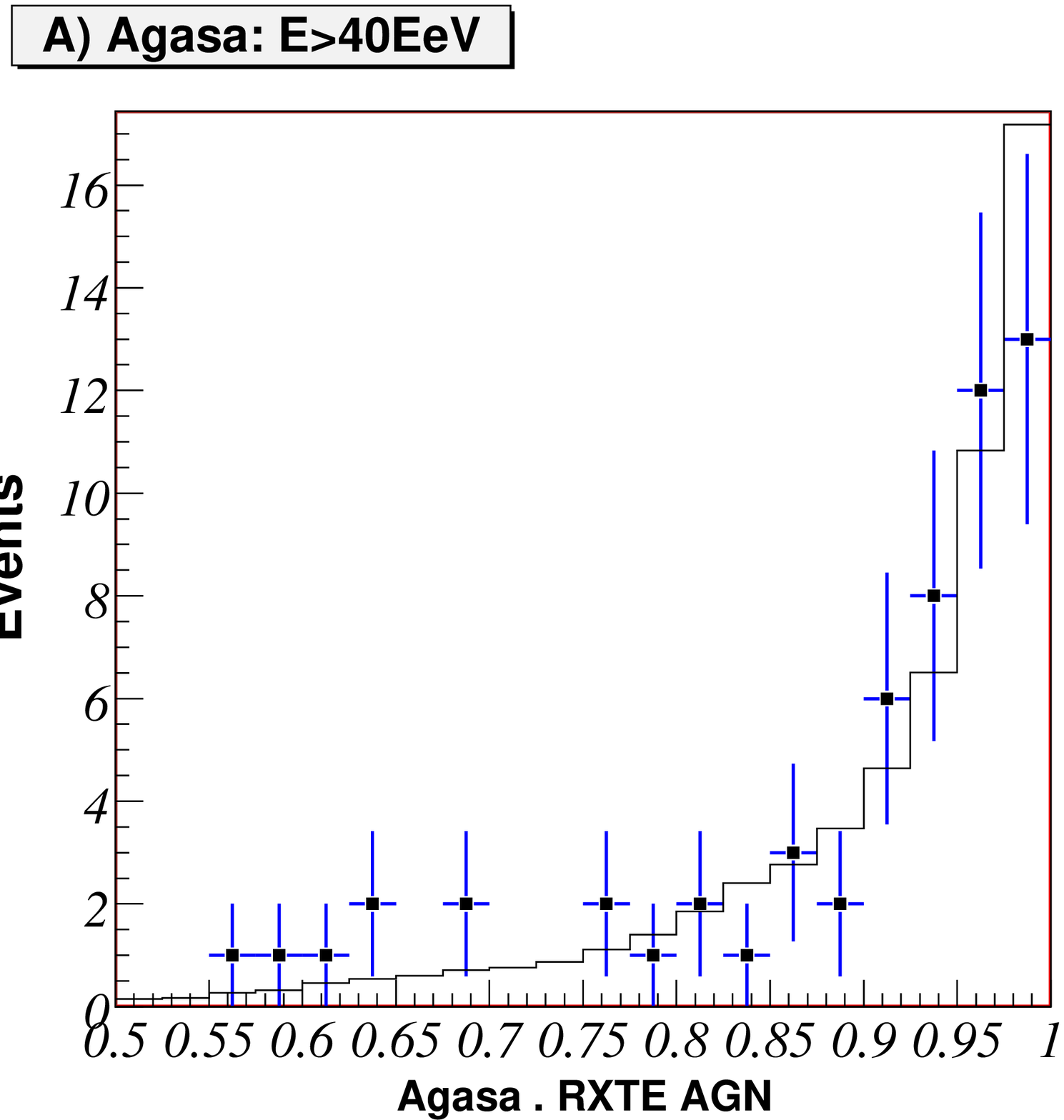} & 
\includegraphics[width=0.5\linewidth]{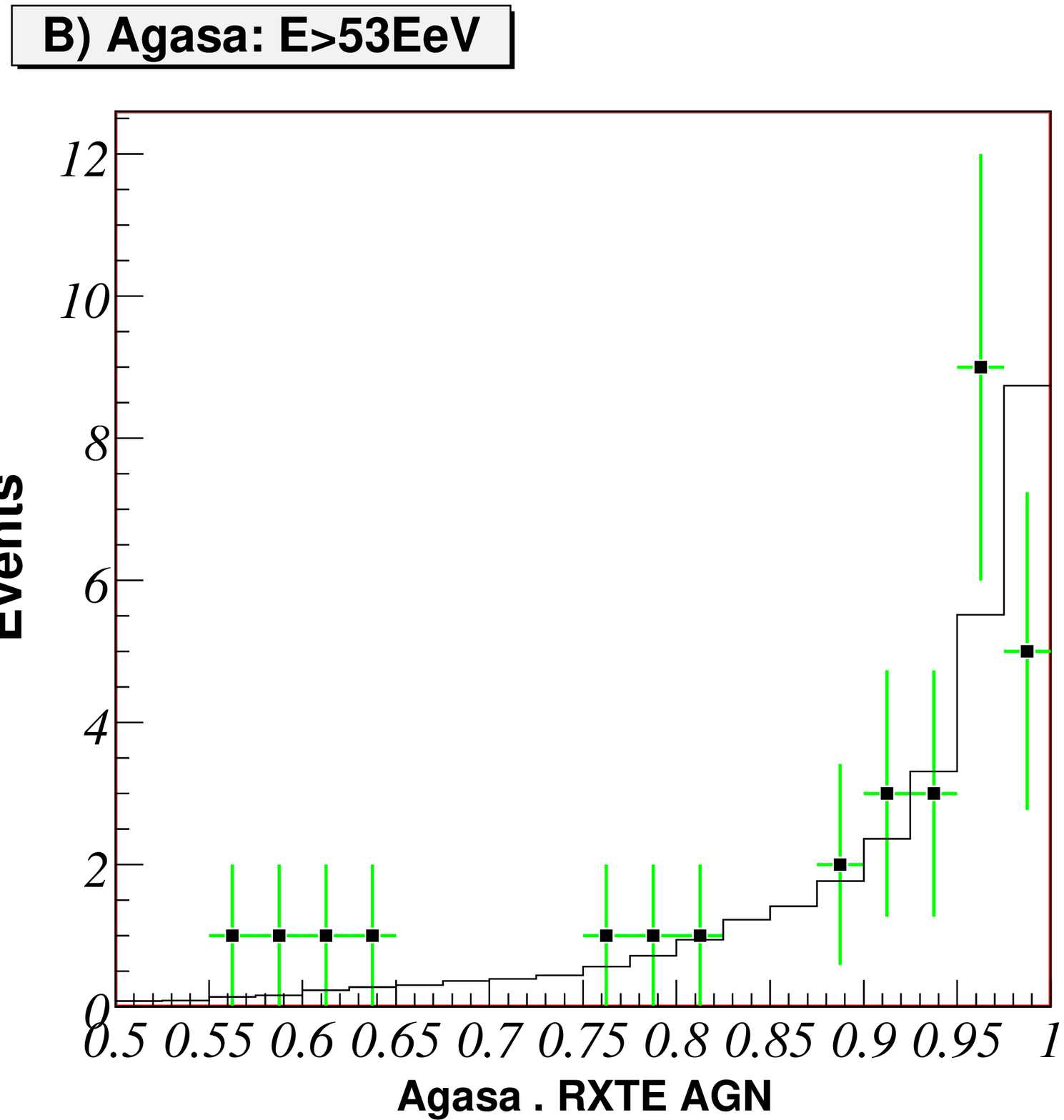}
\end{tabular}
\vspace{0.2cm}
\begin{tabular}{c}
\includegraphics[width=0.5\linewidth]{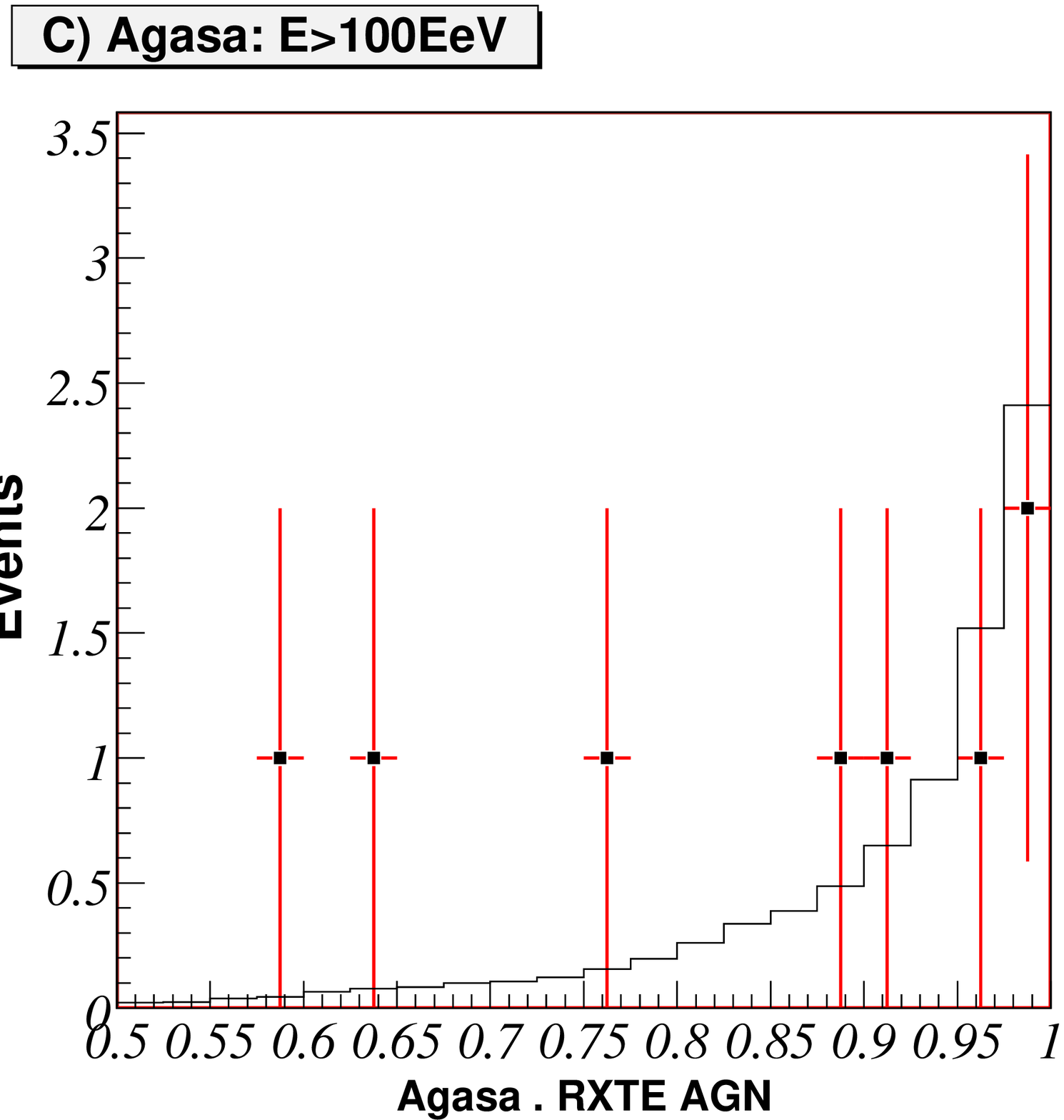}
\end{tabular}
\end{center}
 \caption{{\it \label{fig:broadAGNplot}} The plots show the distribution
 of {\it dot}-products for the {\it broadline}-AGN selection: (top left) with
 cosmic ray energies $E \geq 40$EeV, (top right) with
 cosmic ray energies $E \geq 53$EeV, and (bottom) with
 cosmic ray energies $E \geq 100$EeV.  The curve on each figure shows
 the Monte Carlo {\it random background} normalized
 to the number of entries in each plot. }
\vskip 0.05 cm 
\end{figure}

\begin{figure}[h]
\centering
\includegraphics[width=14cm]{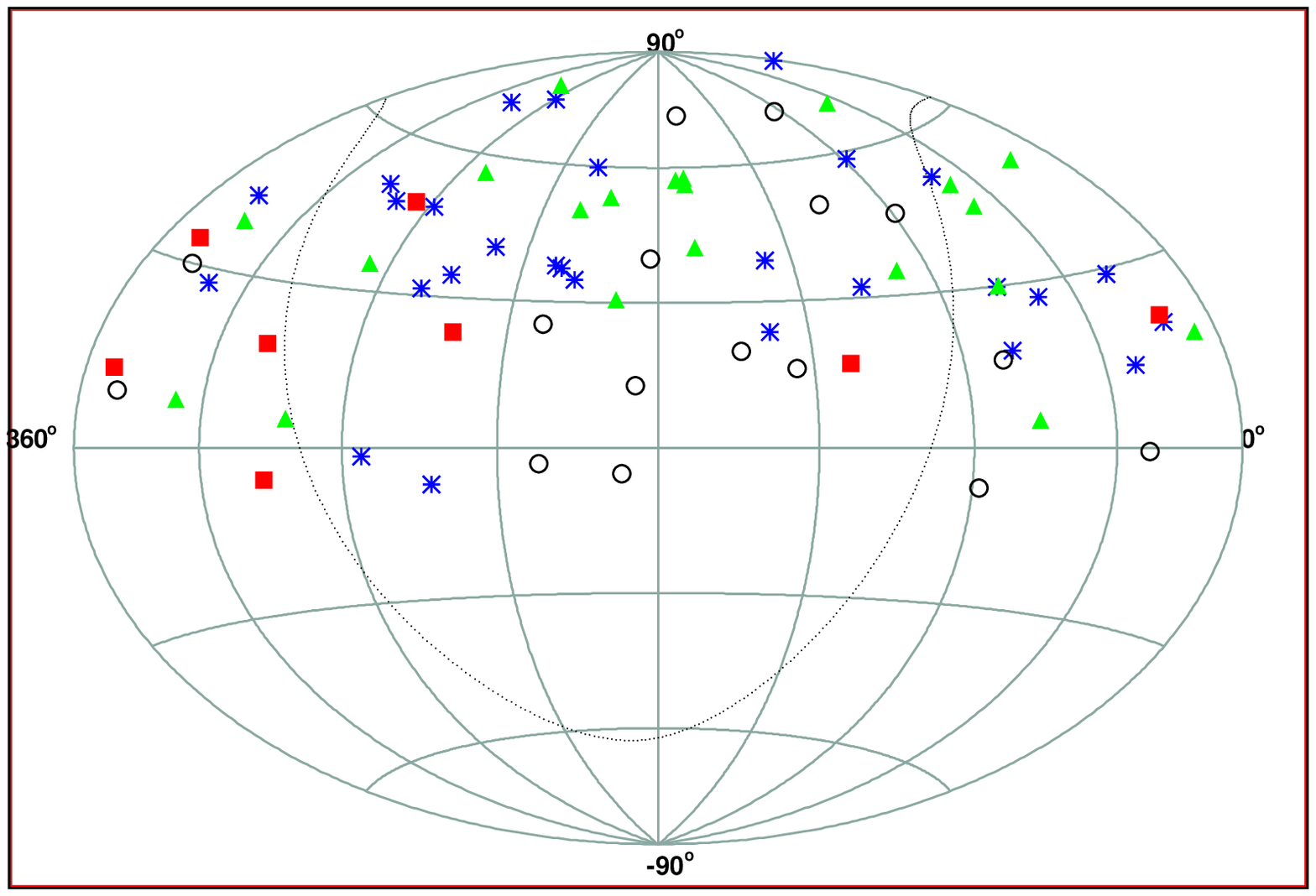}
 \caption{{\it \label{fig:broadAGNdisplay}} The figure shows the 
 map of $RA-Dec$ for the AGASA data and the AGNs from the {\it broadline}- AGN 
selection.  The AGASA data are plotted in
 {\it blue}($*$) for 40EeV$\leq E <$53EeV, 
{\it green}($\blacktriangledown$) for 53EeV$\leq E <$100EeV,
 and  {\it red}($\blacksquare$) 
for 100EeV$\leq E$.  The RXTE AGNs are plotted as {\it black}({\bf o}).  
The galactic plane is drawn as a dotted line. }
\vskip 0.2 cm 
\end{figure}

Plots of the distribution of {\it dot}-products (see definition in text)
for the {\it all}-AGN selection are shown in Fig.\,\ref{fig:allAGNplot}.~
A plot of the AGASA cosmic ray and RXTE AGN directions
are given in Fig.\,\ref{fig:allAGNdisplay}.~
The analogous plots for the 
{\it broadline}-AGN selection are shown
in Fig.\,\ref{fig:broadAGNplot} and \,\ref{fig:broadAGNdisplay}.~
A comparison of 
Figs.\,\ref{fig:allAGNdisplay} and \,\ref{fig:broadAGNdisplay}, 
shows two events shared between the two selections.  Independently we
have verified that all RXTE AGNs with redshift $z\leq0.03$ satisfy at 
least one of our two AGN categories.


The plots of the {\it dot}-products for the 
{\it all}-AGN selection, Fig.\,\ref{fig:allAGNplot},
shows a small excess in the bin nearest to 1 for the AGASA event selections
$E \geq 40$EeV, and $E \geq 53$EeV.  For this bin ({\it i.e.} 
{\it dot}-product $\geq0.975$) the excesses are 
$\sim 1.1$ and $\sim 1.7$ standard deviations for the AGASA event 
selections $E \geq 40$EeV, and $E \geq 53$EeV respectively.  If this
correlation is valid, then it could provide experimental information
to bound the magnetic deflections of extra-galactic cosmic rays.

To see if the {\it all}-AGN category excesses are consistent with {\it e.g.}  
typical GZK models, see Fig.\,\ref{fig:berezinski},
we estimate the RXTE AGN catalog efficiency as follows:
\begin{enumerate}
\item 90\% sky coverage of the $\sim 83$\% of the sky 
surveyed\,\footnote{This corresponds to the sky fraction outside
a $10^{\circ}$ avoidance zone about the galactic plane};
\item 87\% estimated completeness factor;
\item $\sim 77$\% estimated average {\it all}-AGN source detection efficiency 
(from Fig.\,\ref{fig:AGNeffic}).
\end{enumerate}
This yields an lower bound estimate for the 
{\it all}-AGN RXTE efficiency of $\sim 50$\%.
However for the {\it all}-AGN category (only 5 sources, see 
Fig.\,\ref{fig:allAGNdisplay}) it is likely that the global 
(redshift independent) RXTE efficiency factors are not 
appropriate\,\footnote{In particular assuming an average AGN
source density of $4.2 \times 10^{-4}$ Mpc$^{-3}$ (see above) and
the RXTE global efficiency\,\cite{rxte_catalog} of 
$0.9 \times 0.83 \times 0.7 \approx 52$\%, the predicted
number of nearby ($z\leq0.005$) RXTE AGNs is approximated half those observed.
While the small number of AGNs makes this weak statistically, it is
nevertheless consistent with: a RXTE global efficiency of $\sim 100$\% for 
nearby ($z\leq0.005$) AGNs and/or with a local over-density of AGNs.
Although these estimates were based on the AGN number density {\it VS} 
X-ray luminosity from Ref.\,\cite{barger} the AGN number density {\it VS} 
X-ray luminosity deduced by the RXTE experiment\,\cite{rxte_catalog} 
gave a similar result.}.
To obtain an upper bound estimate for the {\it all}-AGN RXTE efficiency 
we assume the global RXTE efficiency is $\sim 100$\%.  Then the
estimated (upper bound) {\it all}-AGN RXTE efficiency is $\sim 77$\%.

The estimated number of cosmic ray:{\it all}-AGN coincidences is: 
the number of cosmic rays (with sources in a given redshift region) 
times the average {\it all}-AGN source detection efficiency 
(for the same redshift region). Thus the estimated number of cosmic rays
from the $z<0.005$ region is obtained by dividing the excess counts,
Fig.\,\ref{fig:allAGNplot}, by the {\it all}-AGN efficiencies 
to obtain: 2.8/0.77 $\sim$ 2.8/0.50 or $3.6 \sim 5.6$ events and
           4.0/0.77 $\sim$ 4.0/0.50 or $5.2 \sim 8.0$ events respectively.  
As fractions of all the observed cosmic rays these are: 
           $3.6 \sim 5.6$/57 (or $6.3 \sim 9.8$\%) and
           $5.2 \sim 8.0$/29 (or $18 \sim 28$\%) respectively.
These fractions are somewhat, to significantly (depending on the {\it all}-AGN 
RXTE efficiency), in excess of typical GZK models 
assuming proton primaries, see Fig.\,\ref{fig:berezinski}.  

Finally we note that the AGASA/HiRes cosmic ray {\it quartet} 
(or possibly {\it quintet}\,\cite{farrar}) cluster, 
at RA $\approx 169.1^{\circ}$, 
Dec $\approx 56.3^{\circ}$\,\cite{hires_cluster}, is near one of the 
the RXTE AGNs at: RA $179.3^{\circ}$, Dec $55.23^{\circ}$
and redshift $z=0.0035$.   In contrast, there is no close 
correlation in the {\it all}-AGN selection with any of the AGASA 
super-GZK events\,\cite{agasa_cluster} plotted in 
Fig.\,\ref{fig:allAGNdisplay}~(more below).

The plots of the {\it dot}-products for the 
{\it broadline}-AGN selection, Fig.\,\ref{fig:broadAGNplot},
are consistent with {\it random background}.
If we assume the cosmic rays are primarily protons, then we can
use a model such as Fig.\,\ref{fig:berezinski} to estimate the number
of cosmic rays expected from sources with redshifts $z\leq0.03$.  Then
{\it e.g.} for the AGASA selection $E \geq 53$EeV, we expect $\sim 60$\%
to originate from sources with redshifts $z\leq0.03$ or $\sim 17.4$ events.
However the number that should appear in {\it dot}-product bins near 1
depends on the RXTE AGN catalog efficiency. Similar to the evaluation
above, we obtain an overall {\it broadline}-AGN RXTE 
efficiency of $\sim 54$\%.
Thus we should observe a {\it signal} as an excess of $\sim 9.4$
events.  Unfortunately in absence of a signal signature ({\it i.e.}
{\it dot}-product bins in excess of {\it random background}) or of
a bound on magnetic field deflections, any statement on lack of
excess depends on the assumed {\it dot}-product range.  That said,
assuming any {\it signal} would appear at {\it dot}-products $\geq0.95$
then $\sim 9.4$ events should result in a $\sim 1.9$ standard deviation excess.
With the AGASA statistics and our current knowledge of cosmic ray
deflections by magnetic fields, no strong conclusion can be drawn.

\begin{figure}[h]
\centering
\includegraphics[width=14cm]{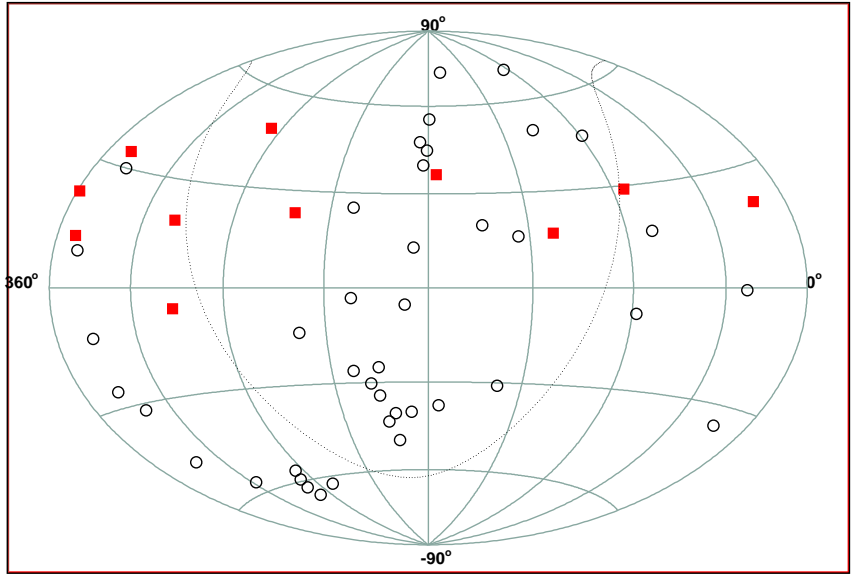}
 \caption{
 {\it \label{fig:allLTp03AGNdisplay}} The figure shows the 
 map of $RA-Dec$ for the updated list\,\cite{AGASAwebsite} 
 AGASA events with energies $E\geq100$EeV and the RXTE catalog 
 of AGNs with redshift $z\leq0.03$ without restriction on AGN declination.  
 The AGASA events are plotted in {\it red}($\blacksquare$).
 The RXTE AGNs are plotted as {\it black}({\bf o}).  
 The galactic plane is drawn as a dotted line.}
\vskip 0.2 cm 
\end{figure}

The final issue is the evidence for, or against, correlations 
between the RXTE catalog of AGNs and the most energetic AGASA events.  
To investigate this, we show in 
Fig.\,\ref{fig:allLTp03AGNdisplay} all of the AGNs from the RXTE catalog
with $z\leq0.03$ and all of the AGASA events above 100EeV as updated on
the AGASA web site\,\cite{AGASAwebsite}.  Now with 11 super-GZK events:
3 have {\it dot}-products $>0.975$, $\sim3$ are close to the galactic
plane (region unobserved by RXTE) and the remaining 5 do not
correlate well ({\it e.g.} {\it dot}-product $^<_{\sim} 0.95$)
with the RXTE catalog of AGNs.  

While the 3 AGASA super-GZK events that are close to RXTE catalog AGNs 
are consistent with {\it random background}, on inspection these events
all have {\it dot}-products $>0.99$.  In this case the expected {\it random 
background} is $\sim 1.6$.  Furthermore one of these events, 
with energy $E=122$EeV and RA $176.0^{\circ}$, Dec $36.3^{\circ}$, is
close to the group of very nearby 
($z<0.005$ in Fig.\,\ref{fig:allAGNdisplay}) AGNs.  
The closest correlation is with the RXTE AGN at: 
RA $182.7^{\circ}$, Dec $39.45^{\circ}$ and redshift $z=0.0033$.

Of the 5 AGASA super-GZK events that do not correlate well with the RXTE 
AGNs, 4 lie far from the galactic plane and have energies well above 100EeV.  
Thus for proton primaries, based on Fig.\,\ref{fig:berezinski}
these should originate at redshifts $z^<_{\sim}0.01$.  For the
{\it broadline}-AGN category the RXTE average 
source detection efficiency is then 
$\sim 100$\% based on Fig.\,\ref{fig:AGNeffic}.  Thus for these AGASA events 
we expect $\sim 4 \times 0.9 \times 0.87 = 3.1$ correlations with the
RXTE catalog of AGNs whereas we observe zero.  Furthermore, the Poisson 
probably of then observing zero is small: 4.4\%.

In contrast if the {\it all}-AGN category is the more appropriate 
source of super-GZK events, then the RXTE average source detection efficiency 
based on Fig.\,\ref{fig:AGNeffic} is significantly less 
than 100\%, particularly for source redshifts to $z^<_{\sim}0.01$.  
In this case, the {\it all}-AGN category the RXTE average source detection 
efficiency is estimated at $\sim 53$\%, resulting in an overall RXTE 
catalog efficiency of $\sim 34$\%.  Thus we expect approximately 
$11 \times 0.34 = 3.7$ correlations (with sources to $z^<_{\sim}0.01$).  The
Poisson probability to observe $\leq 1$ 
correlation (one correlation was observed with the RXTE AGNs 
to $z^<_{\sim}0.01$) is 11.6\%\footnote{If we
also include the 320EeV event from the Fly's Eye\,\cite{FlysEye} then
we expect $12 \times 0.34 = 4.08$ correlations (with RXTE AGNs to
$z^<_{\sim}0.01$) and we observe one.  Now the Poisson probability to
observe $\leq 1$ is 8.6\%.  Anecdotally the Fly's Eye event is very
close, $\sim 3.0^{\circ}$, to one of the RXTE sources at RA $88.8^{\circ}$
Dec $46.3^{\circ}$ and redshift z=0.02. If this
is a true correlation, then the proton nature of the cosmic ray
and/or the measured energy of the cosmic ray are in question.}
However if the AGASA energies are overestimated (with respect to the energy 
scale of Fig.\,\ref{fig:berezinski}) then some of the AGASA events with
energies closest to 100EeV could originate at redshifts $z>0.01$.  
If we extend the possible RXTE AGNs to redshifts of $z^<_{\sim}0.02$ 
then the RXTE average source 
detection efficiency decreases to $\sim 33$\% (because 
most of the lower X-ray luminosity AGNs are unobserved by RXTE) 
resulting in an overall RXTE catalog efficiency $\sim 21$\%.  Thus
we expect $11 \times 0.21 = 2.3$ correlations and we observe two.
The (new) additional correlation is between an AGASA event with $E=120$EeV 
and a RXTE AGN at redshift $z=0.016$.  
The Poisson probability to observe two correlations is $\sim 27$\%.

\section{Summary}
\label{section:summary} 

We have searched for correlations between the published list of
the highest energy events from the AGASA 
experiment\,\cite{agasa_cluster,AGASAwebsite} and the RXTE catalog
of AGNs\,\cite{rxte_catalog}.  Two categories of RXTE AGNs were
considered: {\it all}-AGNs with 
RXTE 3-20keV intrinsic luminosities, $L_{3-20} \geq 10^{41}$ ergs/s,
and {\it broadline}-AGNs with 2-8keV intrinsic luminosities, 
$L_{2-8} \geq 10^{42}$ ergs/s motivated by the analysis of
AGN evolution in Ref.\,\cite{barger}.  To retain RXTE source
detection efficiencies $^>_{\sim}50$\%, source redshifts of
$z\leq0.005$ and $z\leq0.03$ were required for the {\it all}-AGN and
{\it broadline}-AGN categories respectively.

No correlations were observed between the AGASA events and the
{\it broadline}-AGN category of RXTE AGNs even though this category of
AGN is most luminous in X-rays and even though the source density for
this category of AGN is favored by some analyses\,\cite{blasi,kachelriess}
of the highest energy cosmic rays.

In contrast, possible correlations were observed between AGASA events 
and the most inclusive, {\it all}-AGN, 
category of RXTE AGNs.  We note that while 
not statistically conclusive, one of the nearby RXTE AGNs correlates 
with the AGASA/HiRes {\it quartet} event cluster\,\cite{hires_cluster}
and one correlates with one of the AGASA super-GZK events\,\cite{AGASAwebsite}.

Additional data would help confirm, or refute, the 
interesting possibility of highest energy cosmic ray--AGN correlations.

\section{Acknowledgements}

We wish to acknowledge useful communications with Francesc Ferrer
on possible ultra-high energy cosmic rays : BL Lac correlations.

%*******************END OF PLOTTING DATA AND BACKGROUND VARIABLES********

\end{document}